\documentclass[pra,aps,twocolumn,superscriptaddress,nofootinbib,nopacs]{revtex4} 

\usepackage{amsmath}  \usepackage{amssymb}  \usepackage{amsfonts}  \usepackage{bm}  \usepackage{bbm}   \usepackage{braket}  \usepackage{color}  \usepackage{comment}  \usepackage{dcolumn}  \usepackage{enumerate}  \usepackage{epsfig}  \usepackage{gensymb}  \usepackage{graphicx}  \usepackage{indentfirst}  \usepackage{lmodern}  \usepackage{mathrsfs}  \usepackage{mathtools}  \usepackage{psfrag}  \usepackage{pst-all}  \usepackage{soul}  \usepackage{units}  \usepackage{xcolor}
\usepackage{float} 
\usepackage[colorlinks,linkcolor=blue,citecolor=blue,urlcolor=blue,hyperindex,driverfallback=dvipdfm]{hyperref}  \usepackage[T1]{fontenc} 

\def\ii{{\rm i}}  \def\ee{{\rm e}}
\def\me{m_{\rm e}}  
  
\def\Ab{{\bf A}}        \def\Eb{{\bf E}}                              \def\pb{{\bf p}}    \def\qb{{\bf q}}  \def\Rb{{\bf R}}  \def\rb{{\bf r}}    \def\ub{{\bf u}}  \def\vb{{\bf v}} 
\def\xx{\hat{\bf x}}    \def\zz{\hat{\bf z}}            
        
\def\rn{{\bf r}_{\rm n}}  \def\mn{m_{\rm n}}  \def\HH{\hat{\mathcal{H}}}  \def\db{{\bf d}}  \def\zT{z_{\rm T}}

\begin{document} 
\def\bibsection{\section*{\refname}} 

\title{Optical manipulation of matter waves
}

\author{Kamran~Akbari}
\affiliation{ICFO-Institut de Ciencies Fotoniques, The Barcelona Institute of Science and Technology, 08860 Castelldefels (Barcelona), Spain}
\author{Valerio~Di~Giulio}
\affiliation{ICFO-Institut de Ciencies Fotoniques, The Barcelona Institute of Science and Technology, 08860 Castelldefels (Barcelona), Spain}
\author{F.~Javier~Garc\'{\i}a~de~Abajo}
\email{javier.garciadeabajo@nanophotonics.es}
\affiliation{ICFO-Institut de Ciencies Fotoniques, The Barcelona Institute of Science and Technology, 08860 Castelldefels (Barcelona), Spain}
\affiliation{ICREA-Instituci\'o Catalana de Recerca i Estudis Avan\c{c}ats, Passeig Llu\'{\i}s Companys 23, 08010 Barcelona, Spain}


\begin{abstract}
Light is extensively used to steer the motion of atoms in free space, enabling cooling and trapping of matter waves through ponderomotive forces and Doppler-mediated photon scattering. Likewise, light interaction with free electrons has recently emerged as a versatile approach to modulate the electron wave function for applications in ultrafast electron microscopy. Here, we combine these two worlds by theoretically demonstrating that matter waves can be optically manipulated via inelastic interactions with optical fields, allowing us to modulate the translational wave function and produce temporally and spatially compressed atomic beam pulses. Specifically, we realize such modulation through stimulated photon absorption and emission by atoms traversing phase-matching evanescent optical fields generated upon light scattering by a nanostructure, but also via stimulated Compton scattering in free space without any assistance from material media. Our results support optical manipulation of matter waves as a powerful tool for microscopy, spectroscopy, and the exploration of novel fundamental phenomena associated with light-atom interactions.
\end{abstract}

\maketitle 

\section{Introduction}

Light enables exquisite control over the excitation dynamics of atoms \cite{DHK02}, molecules \cite{WPL16}, and materials \cite{MYH20} in general through photon exchanges with internal electronic and vibrational degrees of freedom. In addition, light can directly act on the translational wave function of matter waves to steer their dynamics \cite{GRP1986,BSB1988,BLS1989,ASM94,CSP09}, while optical trapping of objects ranging in size from atoms to human cells has become commonplace both in cold atom physics \cite{L01} and in the manipulation of living microorganisms \cite{AD1987,paper140}.

In parallel, the interaction between free electrons and optical fields has recently been recognized as a valuable tool to modulate the wave function of the former, a capability that is emphasized in the so-called photon-induced near-field electron microscopy (PINEM) \cite{BFZ09,paper151,PLZ10,FES15,PLQ15,EFS16,KSE16,RB16,VFZ16,paper272,paper282,PRY17,KML17,FBR17,paper311,paper312,paper306,paper325,paper332,K19,PG19,PZG19,paper339,DNS20,KLS20,WDS20,VMC20,KDS21,ZSF21,HRF22}. In this technique, electron-light coupling materializes in the direct absorption and emission of individual photons by the moving electron, subject to the condition that the electron velocity $v$ matches the phase velocity $\omega/q$ associated with some light components of frequency $\omega$ and wave vector $q$ along the direction of particle motion. Obviously, this requires evanescent fields (i.e., $q>\omega/c$) such as those generated by light scattering at nanostructures \cite{H09,paper114}. By capitalizing on the availability of ultrashort laser pulses, PINEM allows for the mapping of material excitations and confined optical fields to be performed with combined space-time resolution in the nanometer-femtosecond domain \cite{PLQ15,paper282,paper306,KDS21}. In addition, light-electron interaction has been demonstrated to transfer linear \cite{paper311,FYS20} and angular \cite{paper332,paper312} momenta and thus shape the transverse profile of the electron wave function, while also enabling temporal compression of the longitudinal wave function down to the attosecond regime \cite{FES15,PRY17,MB18_2}.

Free-electron modulation in empty space is possible as well through ponderomotive interaction with optical fields, leading to stimulated Compton scattering \cite{KES17} and Kapitza-Dirac electron diffraction \cite{KD1933,FAB01,FB02,B07,TL19,ACS20}, which also enable temporal compression \cite{BZ07,KSH18} and lateral phase imprinting \cite{SAC19}, with potential application in actively controlled focusing and steering of electron beams \cite{MJD10,paper368}. As a practical consideration, while PINEM interaction is proportional to terms in the light-matter coupling Hamiltonian that scale linearly with the optical vector potential $\Ab$, ponderomotive effects involve $A^2$ terms that remain much weaker up to light intensities as high as \cite{paper371} $\sim10^{19}\,$W/cm$^2$ for a typical electron energy of 100\,keV.


We expect that optical control over the translational motion of neutral atoms and molecules can also be exerted in a similar way as for free electrons, with the advantage that coupling to light becomes stronger when it involves polarization of the internal electronic degrees of freedom. Beyond its fundamental interest, this type of interaction could find practical application in the creation of temporally compressed atom waves, in analogy to electron pulse compression in PINEM. In addition, we anticipate a new form of photon-induced near-field atom microscopy (PINAM), in which atoms instead of electrons are used to probe localized excitations, providing strong interaction with light.

\begin{figure*}
\centering
\includegraphics[width=1.0\textwidth]{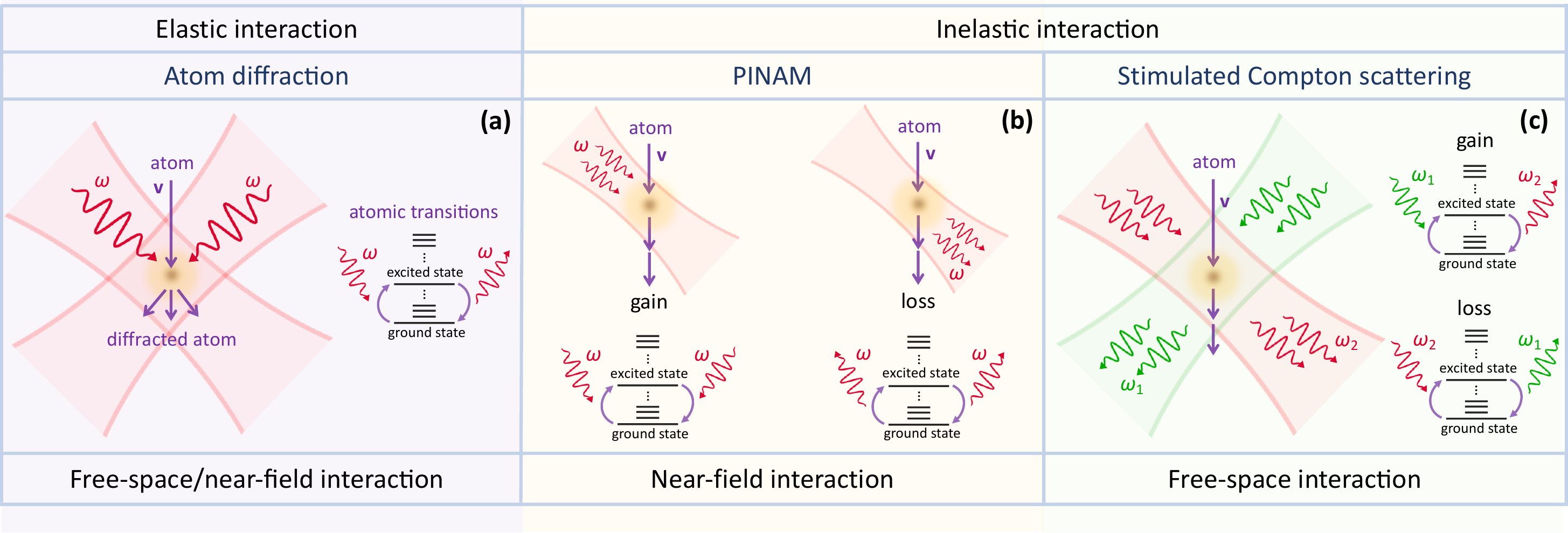}
\caption{{\bf Free-atom interactions with optical fields.} ({\bf a})~Elastic scattering by spatially modulated monochromatic light can change the direction of the atom CM momentum, thus producing diffraction of matter waves \cite{KD1933,FAB01,FB02,B07,TL19,ACS20,SAC19}. ({\bf b})~Direct photon absorption or emission is possible under illumination with monochromatic evanescent fields containing wave vector $\qb$ and frequency $\omega$ components satisfying the phase-matching condition $\qb\cdot\vb=\omega$, where $\vb$ is the atom velocity. Two photons are involved in each event to leave the atom in its original ground state, so that the energy transfer is entirely associated with the CM motion, opening the way to perform photon-induced near-field atom microscopy (PINAM). Energy gain and loss processes are both enabled, as indicated in the lower sketches. ({\bf c})~Stimulated Compton scattering (SCS) under illumination by two light plane waves (green and red, $i=1,2$) of wave vectors $\qb_i$ and frequencies $\omega_i$ results in a net change in the CM energy by $\hbar(\omega_1-\omega_2)$ (green to red, atom energy gain) or $-\hbar(\omega_1-\omega_2)$ (red to green, energy loss), provided the phase-matching condition $\omega_1-\omega_2=(\qb_1-\qb_2)\cdot\vb$ is satisfied. Electronic transitions mediate these processes, in which internal excitation is followed by de-excitation to leave the atom in the ground state.}
\label{Fig1}
\end{figure*}

Here, we theoretically demonstrate that matter waves can be manipulated by stimulated photon scattering in the presence of optical fields, enabling modulation of the translational wave function of atomic beams and opening the way to unprecedented spatial compression. Although elastic photon scattering is well-known to produce changes in the center-of-mass (CM) momentum, as well as associated diffraction effects \cite{KD1933,FAB01,FB02,B07,TL19,ACS20,SAC19} (Fig.~\ref{Fig1}a), we are instead interested in inelastic interactions, which we investigate within two conceptually different types of scenarios: ($i$) direct photon absorption or emission by an atom moving in the presence of evanescent fields (Fig.~\ref{Fig1}b); and ($ii$) stimulated Compton scattering (SCS) under two-color illumination in free space (Fig.~\ref{Fig1}c). The studied inelastic processes affect the CM of the atom, while the photon energies are taken to differ from the internal electronic resonances, such that the atom is left in its original ground state after the interaction has taken place. Substantial light-atom coupling is then achieved for relatively moderate optical fields, provided some rigorously established phase-matching conditions are satisfied, leading to the emergence of coherent sidebands in the CM energy (i.e., belonging to the same wave function), and enabling spatial compression of the CM wave function upon free-space propagation. We anticipate applications in the production of extreme localization of free-space atoms, while new states of matter could also be explored as an extension of these ideas to atom bunches.

\begin{figure*}
\centering
\includegraphics[width=0.9\textwidth]{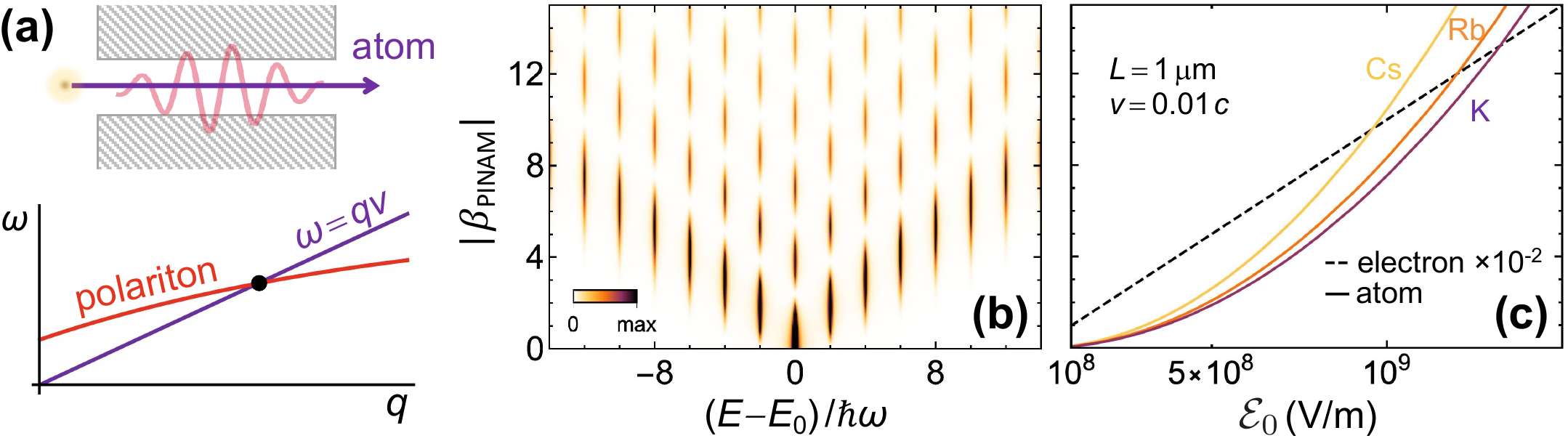}
\caption{{\bf PINAM interaction.} ({\bf a})~Monochromatic illumination with evanescent fields of well-defined wave vector $q$ (along the atom beam direction) and frequency $\omega$ can be implemented for atoms traversing a hole drilled in a polaritonic material (upper  sketch) when the external light excites hole-confined polaritons under phase-matching conditions (indicated by the black dot in the dispersion relation). ({\bf b})~An atom incident with energy $E_0$ develops sidebands in the CM spectrum at energies $E_\ell=E_0+2\ell\hbar\omega$ with probability $J_\ell^2(2|\beta_{\rm PINAM}|)$ only dependent on a light-atom PINAM coupling coefficient $\beta_{\rm PINAM}$. The atom CM spectrum evolves with $\beta_{\rm PINAM}$ in a similar way as in PINEM, but with sidebands separated by $2\hbar\omega$ instead of $\hbar\omega$. Sidebands are convoluted here with a Lorentzian of $0.12\,\hbar\omega$ FWHM and the spectral integral over $E$ is normalized to unity. ({\bf c})~The coupling coefficient $\beta_{\rm PINAM}$ scales as $L\,\mathcal{E}_0^2/v$ with the hole length $L$, the atom velocity $v$, and the optical electric field amplitude $\mathcal{E}_0$. We represent $|\beta_{\rm PINAM}|$ as a function of $\mathcal{E}_0$ for different atoms moving with velocity $v=0.01\,c$ along a path length $L=1\,\mu$m (solid curves). For comparison, we show the PINEM coupling coefficient $\beta_{\rm PINEM}\propto L\,\mathcal{E}_0$ for electrons moving under the same conditions and a photon energy $\hbar\omega=1\,$eV (dashed line).}
\label{Fig2}
\end{figure*}

\section{Inelastic interaction between free atoms and optical fields}

We based our results on a rigorous quantum description of free atoms, which leads to an effective Hamiltonian for the CM motion, encapsulating the internal atomic degrees of freedom and accounting for both elastic and inelastic atom-photon processes. The object of our study is a neutral atom moving with nonrelativistic velocity $\vb$ in the presence of external illumination, which is introduced through its associated classical vector potential $\Ab(\rb,t)$ within the minimal coupling scheme in the Coulomb gauge. We assume moderate light intensities and optical frequencies that are far from any electronic resonance compared with the width of the latter, so that the atom is just slightly perturbed from its initial internal ground state during the interaction period. Then, starting from a complete quantum description of the moving atom \cite{M1966,BJ03} and assuming Markovian dynamics \cite{L1973}, the electronic degrees of freedom can indeed be absorbed in an effective Hamiltonian $H^{\rm eff}(\rb,t)$ that describes the time- and position-dependent atomic CM wave function $\psi(\rb,t)$. More precisely, we obtain the Schr\"odinger equation (see Appendix)
\begin{align}
\ii\hbar\,\dot{\psi}(\rb,t)\!=\!\!\bigg[\!E_0\!-\hbar(\ii\nabla_\rb+\qb_0)\!\cdot\!\vb+H^{\rm eff}(\rb,t)\!\bigg]\psi(\rb,t)
\label{schr2}
\end{align}
with
\begin{align}
&H^{\rm eff}(\rb,t)=\frac{e^2}{\hbar c^2} \sum_{j\neq0} \omega_{j0}\,|\xx\cdot\db_{j0}|^2 \;\Ab(\rb,t)\cdot\tilde{\Ab}_j(\rb,t), \label{Heffsimplified}
\end{align}
where $j$ runs over excited electronic states, $\omega_{j0}$ and $e\db_{j0}$ are the frequency and dipole moment associated with a transition between the ground state $\ket{0}$ and $\ket{j}$, and
\begin{align}
&\tilde{\Ab}_j(\rb,t)=\!-\ii\omega_{j0}\!\int_{-\infty}^t\!\!\!dt\,\ee^{-\ii\omega_{j0}(t-t')}
\,\Ab(\rb-\vb t+\vb t',t') \nonumber
\end{align}
is a $j$-dependent modified vector potential. These expressions are applicable to isotropic atoms and relatively small CM deflections, such that the wave function only contains energy and momentum components that are tightly packed around central values $E_0$ and $\hbar\qb_0$ (nonrecoil approximation \cite{paper371}, see Appendix).

Assuming an optical field acting over a finite interaction interval, the solution to Eq.~\eqref{schr2} admits the form
\begin{align}
\psi(\rb,t)=\psi^{\rm inc}(\rb,t)\,\ee^{-(\ii/\hbar)\int_{-\infty}^t \!\!dt'\, H^{\rm eff}(\rb-\vb t+\vb t',t')},
\label{solution}
\end{align}
where $\psi^{\rm inc}(\rb,t)$ is the incident wave function in the absence of external illumination. Implicit in this result is the assumption that the nonrecoil approximation remains valid along the interaction region (i.e., the atom beam is well collimated and such region is small enough as to dismiss lateral diffraction). For propagation over large distances beyond the interaction interval, diffraction can be readily incorporated through free-space propagation according to the remaining Hamiltonian $-\hbar^2\nabla_\rb^2/2M$, where $M$ is the atomic mass.

In what follows, we explore solutions corresponding to the aforementioned PINAM and SCS scenarios, for which the post-interaction wave function takes the general form
\begin{align}
&\psi(\rb,t)=\psi^{\rm inc}(\rb,t)\,\ee^{\ii\varphi(\Rb)} \label{psigeneral}\\
&\times\!\!\!\sum_{\ell=-\infty}^\infty \!\!J_\ell(2|\beta(\Rb)|)\,\ee^{\ii\ell {\rm arg}\{-\beta(\Rb)\}}\,\ee^{\ii\ell(z-vt)\Omega/v}\,\ee^{-2\pi\ii\ell^2d/\zT}, \nonumber
\end{align}
where $\beta(\Rb)$ is a coupling coefficient that depends on the coordinates $\Rb$ in a plane perpendicular to $\vb$, $\varphi(\Rb)$ is an elastic phase, $\Omega$ is an optical frequency related to the incident photon characteristics, and the $\ell$ sum describes periodically spaced energy sidebands separated by $\hbar\Omega$ with associated shifts in longitudinal momentum by multiples of $\hbar\Omega/v$. We have incorporated in Eq.~\eqref{psigeneral} the effect of velocity spreading acting over a long distance $d$ after interaction. This involves a Talbot distance \cite{paper360} $\zT=4\pi Mv^3/\hbar\Omega^2$, where we neglect relativistic corrections. We remark that this solution not only includes diffraction by the spatial texture of the optical field intensity through the phase $\varphi(\Rb)$, but also kinetic energy jumps resulting from events of direct photon absorption or emission by the atom CM motion, as determined by the coupling coefficient $\beta(\Rb)$, which is the main focus of this work.

\begin{figure*}
\centering
\includegraphics[width=0.9\textwidth]{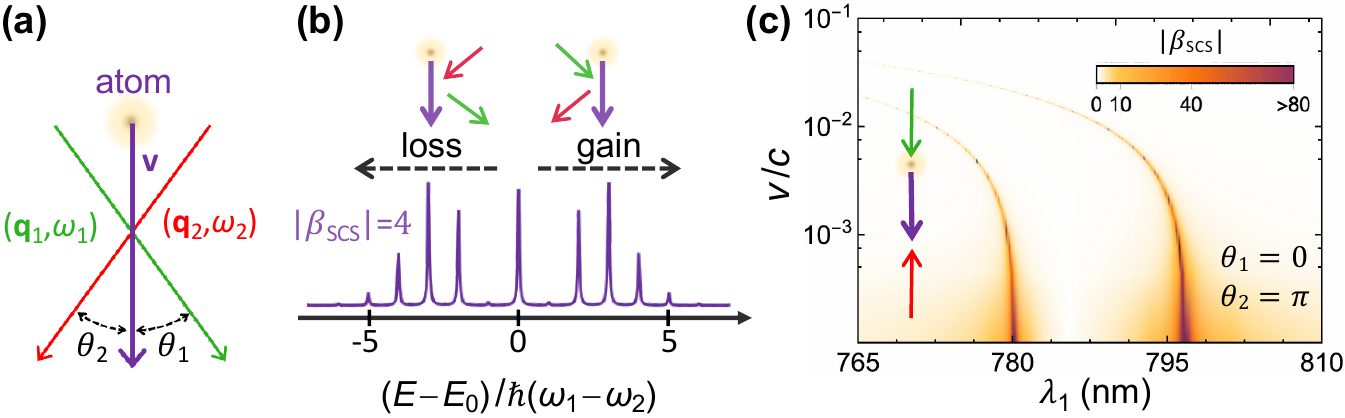}
\caption{{\bf Free-space optical manipulation of matter waves.} ({\bf a})~Stimulated Compton scattering can be achieved in free space by irradiation with two light plane waves of frequencies and angles satisfying the phase-matching condition $\omega_1/\omega_2=(c-v\cos\theta_2)/(c-v\cos\theta_1)$. ({\bf b})~Atom CM spectra similar to those in Fig.~\ref{Fig2}b are then generated, with sidebands separated by the photon energy difference $\hbar(\omega_1-\omega_2)$, as illustrated here for a coupling coefficient $|\beta_{\rm SCS}|=4$. Energy gains and losses of the CM are associated with absorption and emission events as indicated in the upper inset. ({\bf c})~We plot the coupling coefficient $|\beta_{\rm SCS}|$ for a Rb atom under illumination with angles $\theta_1=0$ and $\theta_2=\pi$ (see inset) as a function of atom velocity and the light wavelength $\lambda_1=2\pi c/\omega_1$ near the 5s-5p optical transition. The other light wavelength $\lambda_2$ is determined by the above phase-matching condition. We take $L=1\,$mm for the length of the interaction region and $|\Eb_1\cdot\Eb_2^*|=1\,$(MV/m)$^2$ for the product of the two optical light fields.}
\label{Fig3}
\end{figure*}

\section{PINAM interaction}

A direct consequence of Eq.~\eqref{solution} is that direct absorption or emission of optical quanta by the moving atom is only possible for field components of wave vector $\qb$ and frequency $\omega$ satisfying the phase-matching condition
\begin{align}
\omega=\qb\cdot\vb, \label{phasematching1}
\end{align}
(i.e., evanescent fields \cite{paper371}). Slow polaritons are well-suited to satisfy this condition, which is illustrated in Fig.~\ref{Fig2}a by the point of crossing between their dispersion relation and the electron line. As a practical configuration, we consider a polaritonic material drilled with holes in which one-dimensional modes are confined and excited through external illumination. Atoms that traverse the film along the holes see their wave function modified as prescribed by Eq.~\eqref{psigeneral} with $\Omega=2\omega$ and a coupling coefficient (see Appendix)
\begin{align}
|\beta|=|\beta_{\rm PINAM}|\equiv\frac{L}{2\hbar v}\,\alpha_0|\mathcal{E}_0|^2, \label{betamaintext}
\end{align}
where $L$ is the film thickness, $\mathcal{E}_0$ is the electric field amplitude associated with the polariton, and $\alpha_0$ is the static atomic polarizability. Each sideband $\ell$ in Eq.~\eqref{psigeneral} has a probability of occupation given by $J_\ell^2(2|\beta_{\rm PINAM}|)$, which produces a pattern of atom CM energy distributions as a function of $|\beta_{\rm PINAM}|$ as illustrated in Fig.~\ref{Fig2}b, similar to those observed in PINEM \cite{paper151,FES15}, but with sidebands separated by an energy spacing $\hbar\Omega=2\hbar\omega$, instead of just $\hbar\omega$. This reflects the fact that the neutral atom necessitates two photons to realize a PINAM interaction (one to excite and another one to de-excite the internal degrees of freedom, see Fig.~\ref{Fig1}b). We show in Fig.~\ref{Fig2}c the values of $|\beta_{\rm PINAM}|$ obtained for selected alkali atoms (K, Rb, and Cs, for which $\alpha_0=42.9$, $47.4$, and $59.4\,{\AA}^3$, respectively \cite{SN19}), which present relatively high static polarizabilities. Sizeable values of the coupling coefficient are thus expected for propagation over a distance of $1\,\mu$m and attainable field amplitudes compatible with material damage during sub-ps illumination times. Incidentally, assuming the same phase-matching conditions, the PINEM coupling coefficient for electrons is \cite{paper371} $|\beta_{\rm PINEM}|=eL|\mathcal{E}_0|/\hbar\omega$, independent of velocity, linear in the field amplitude, and exceeding the PINAM coefficient by several orders of magnitude for a photon energy $\hbar\omega=1\,$eV (see Fig.~\ref{Fig2}c).

\section{Near-resonant stimulated Compton scattering}

A disadvantage of PINAM is that it requires a material structure to mediate the atom-light interaction, and this imposes severe restrictions on the optical modes and light intensities that can be employed, as well as a minimum atom energy required to produce phase matching. We therefore investigate an alternative in which coherent energy sidebands can be produced in free space via SCS under illumination with two light plane waves ($i=1,2$), as illustrated in Fig.~\ref{Fig3}a. In this configuration, a photon is taken from one of the beams and scattered along the other beam, thus rendering a stimulated process with a probability proportional to the product of the two incident beam intensities. We consider $v\ll c$ as a typical condition that is additionally beneficial to reach strong light-atom coupling (see below), although the present study can be straightforwardly extended to higher atom velocities. Defining the associated electric field amplitudes $\Eb_i$, wave vectors $\qb_i$, and photon frequencies $\omega_i$, the CM wave function after interacting over a finite path length $L$ is again given by Eq.~\eqref{psigeneral} with $\Omega=\omega_1-\omega_2$ and a coupling coefficient (see Appendix)
\begin{align}
\beta\!=\!\beta_{\rm SCS}\!\equiv\!-\frac{2\ii e^2\omega_1L}{\hbar^2v\,\omega_2}\Eb_1\!\cdot\Eb_2^*\,
\!\sum_{j\neq0}\frac{\omega_{j0}\,|\xx\cdot \db_{0j}|^2}{\omega_{j0}^2\!-(\omega_1\!-\!\qb_1\!\cdot\!\vb)^2},
\label{betaCompton}
\end{align}
subject to the phase-matching condition
\begin{align}
\omega_1-\omega_2=(\qb_1-\qb_2)\cdot\vb. \label{phasematching2}
\end{align}
The resulting CM wave function is again similar to PINEM \cite{paper371}, but composed of sidebands corresponding to energy transfers that are multiples of the difference between the two incident photon energies, $\hbar(\omega_1-\omega_2)$ (Fig.~\ref{Fig3}b).

An advantage of this scheme with respect to PINAM is that the interaction can be dramatically enhanced by tuning the Doppler-shifted frequency $\omega_1-\qb_1\cdot\vb=\omega_2-\qb_2\cdot\vb$ close to one of the atomic resonances $j$. This is illustrated in Fig.~\ref{Fig3}c, where $|\beta_{\rm SCS}|$ is plotted as obtained from Eq.~\eqref{betaCompton} for Rb atoms near the 5s-5p optical transition region, dominated by two resonances ($j=1,2$) with parameters \cite{SWC03} $\hbar\omega_{10}=1.56\,$eV, $\hbar\omega_{20}=1.59\,$eV, $|\xx\cdot\db_{10}|^2=1.75\,{\AA}^2$, and $|\xx\cdot\db_{20}|^2=3.30\,{\AA}^2$. We ignore the resonance finite widths ($\Delta \lambda<10^{-5}\,$nm) under the assumption that the optical frequencies are at least several widths apart from the atomic resonances.
Specifically, we consider two counter-propagating light beams at angles $\theta_1=0$ and $\theta_2=\pi$ relative to the atom velocity $\vb$, which maximize the transition frequency $\omega_1-\omega_2=[2v/(c+v)]\,\omega_1$ obtained from Eq.~\eqref{phasematching2} and further provide the additional advantage of enabling illumination over a long atom path length $L$, such that high values of $|\beta_{\rm SCS}|$ can be achieved using moderate light fields. Indeed, we find $|\beta_{\rm SCS}|\gg1$ near the atomic resonances for optical field amplitudes of 1\,MV/m (i.e., an intensity of $\sim0.5\,$MW/cm$^2$) acting along 1\,mm (see Fig.~\ref{Fig3}c). Incidentally, a blue shift is observed in the resonance conditions as $v$ increases, while the $\beta_{\rm SCS}\propto1/v$ scaling makes it easier to reach large coupling at low velocities for a detuning of a few nm in the incident light wavelengths.


\begin{figure}
\centering
\includegraphics[width=0.45\textwidth]{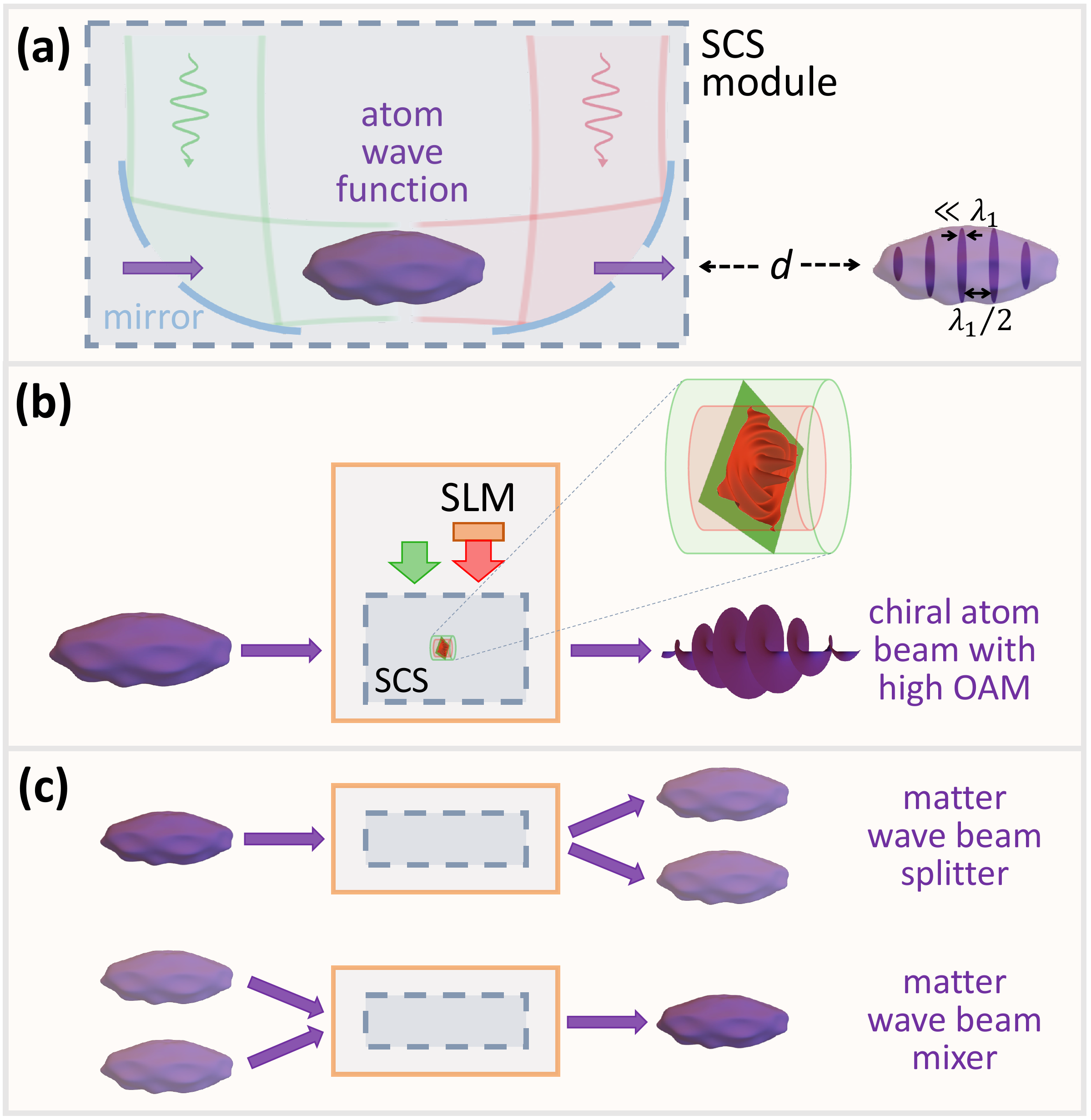}
\caption{{\bf Tools for optical manipulation of matter waves.} ({\bf a})~Following SCS interaction under the conditions of Fig.~\ref{Fig3}c (see mirror-based scheme for light injection), temporal compression of the CM wave function is achieved after free-space propagation over a distance $d$. A train of atom probability density pulses is thus generated, separated by approximately half of the light wavelength $\lambda_1/2$. ({\bf b})~By laterally shaping one of the light beams with a spatial light modulator (SLM), its phase is directly imprinted on the inelastic components of the atom wave function to produce designated profiles (e.g., a chiral beam in the scheme). ({\bf c})~Beam splitting and mixing can be performed by patterning an optical grating through the SLM.}
\label{Fig4}
\end{figure}

\section{Concluding remarks}

The atom-light inelastic interactions here explored open a viable approach to shape the translational wave function of free-space atoms boldly beyond currently existing capabilities \cite{GRP1986,BSB1988,BLS1989,ASM94,DGR99,SF02,OKS05}. In particular, the emergence of coherent sidebands in the kinetic energy distribution leads to temporal compression upon propagation over a large distance $d$ after the interaction has taken place (Fig.~\ref{Fig4}a), as demonstrated in the context of electron accelerators \cite{SCI08}, and later using smaller setups to achieve attosecond compression by exploiting ponderomotive \cite{BZ07,KSH18} and PINEM \cite{FES15,PRY17,MB18_2} interactions. Importantly, a high degree of compression approaching the classical point-particle limit is attainable by combining several cycles of optical interaction spaced by free propagation \cite{YFR21}. For a coupling strength $|\beta|\gg1$, maximum compression is observed in the CM wave function of Eq.~\eqref{psigeneral} after a propagation distance \cite{ZSF21,paper373} $d\approx0.0733\,\zT/|\beta|$. For example, for Rb atoms (mass $M\approx85.5\,$u) and $|\beta|=10$, PINAM interaction under the conditions of Fig.~\ref{Fig2}c leads to $d\sim40\,$cm, whereas we have $d\propto v$ in the SCS configuration of Fig.~\ref{Fig3}c (i.e., $\sim800\,$nm light wavelength), with $d\sim5\,$mm for a velocity $v\sim10^5\,$m/s. Temporal compression of atom waves opens a range of possible applications in precise atom implantation \cite{R07} and controlled atom collisions.


Control over the transverse wave function is also possible through the phase of both $\varphi$ and $\beta$. The former is commonly exploited in atom optics and scales linearly with the field intensity, whereas the phase in $\beta$ for SCS is directly given by the differences between the phases of the two externally applied light waves [see Eq.~\eqref{betaCompton}], thus permitting us to directly imprint an on-demand lateral wave function and prepare, for example, states with a desired degree of orbital angular momentum \cite{paper332,paper312} (Fig.~\ref{Fig4}b) by optical interactions instead of mask patterning \cite{LSD21}, as well as focused atomic beams \cite{SPS10} with tailored profiles \cite{paper368}. This approach could be used to implement beam splitters and mixers (Fig.~\ref{Fig4}c), and further combined to assemble interferometers \cite{CSP09}, quantum erasers, and delayed-choice setups \cite{MKZ16}.

The present results can be readily extended to molecular beams, thus suggesting applications in precise epitaxial growth, as well as in the control of nanoscale chemical reactions. A direct extension of the PINEM formalism to ions shows that the coupling coefficient should be linear in the charge. However, free-space optical manipulation of ions through ponderomotive forces is limited by the inverse-mass dependence of the $A^2$ coupling Hamiltonian. In contrast, the coupling coefficient is independent of the atomic mass in the SCS configuration here investigated for neutral atoms.

By incorporating resonant components in the external illumination, additional means of control of atoms and molecules are enabled by engaging their internal excitations, which could be activated, for example, through $\pi$ pulses introduced at designated positions along the particle path, thus reconfiguring the set of available optical transitions to engineer the SCS interaction, and also providing an alternative source of CM momentum. In molecules with rotational and vibrational degrees of freedom, low-energy excitations provide an extra range of resonance frequencies, over two orders of magnitude below electronic excitations, thus suggesting the use of external illumination extending from the infrared to the microwaves for near-resonant inelastic SCS manipulation.

\acknowledgments

This work has been supported in part by the European Research Council (Advanced Grant 789104-eNANO), the European Commission (Horizon 2020 Grants 101017720 FET-Proactive EBEAM and 964591-SMART-electron), the Spanish MICINN (PID2020-112625GB-I00 and Severo Ochoa CEX2019-000910-S), the Catalan CERCA Program, and Fundaci\'{o}s Cellex and Mir-Puig.


\appendix 

\section{Light-atom interaction: transition matrix elements}

The free-atom Hamiltonian $\HH$ admits a complete, orthonormal set of eigenstates $\ket{\pb j}$ (i.e., $\sum_{\pb j}\ket{\pb j}\bra{\pb j}=\mathcal{I}$ and $\bra{\pb j}\pb'j'\rangle=\delta_{\pb\pb'}\delta_{jj'}$) factorized into CM and internal components \cite{M1966,BJ03}, which we label by the total translational momentum $\hbar\pb$ and an internal state index $j$, respectively. We can thus write $\HH^\mathrm{at}\ket{\pb j}=\hbar(\varepsilon_\pb+\omega_j)\ket{\pb j}$, where $\hbar\omega_j$ is the internal energy, $\hbar\varepsilon_\pb=\hbar^2p^2/2M$ is the CM energy, and $M$ is the atomic mass. In what follows, we denote the nuclear charge, mass, and position as $eZ$, $\mn$, and $\rn$, respectively, while the bound electron coordinates $\rb_i$ are labelled by an index $i=1,\dots,Z$ and the electron charge and mass are $-e$ and $\me$. We also use the CM coordinates
\begin{subequations}
\label{rbub}
\begin{align}
\rb=\frac{1}{M}\bigg(\mn\rn+\me\sum_i\rb_i\bigg),
\end{align}
the electronic positions
\begin{align}
\ub_i=\rb_i-\rn
\end{align}
\end{subequations}
relative to the nucleus, and the nucleus position $\ub=\rn-\rb=-(\me/M)\sum_i\ub_i$ relative to the center of mass.

The total Hamiltonian of the system $\HH(t)=\HH^\mathrm{at}+\HH'(t)+\HH''(t)$ incorporates coupling to classical radiation through the terms
\begin{widetext}
\begin{subequations}
\label{Hint12}
\begin{align}
\HH'(t)&\!=\!\frac{\ii\hbar eZ}{\mn c}\Ab(\rn,t)\!\cdot\!\nabla_{\rn}\!-\!\frac{\ii\hbar e}{\me c}\sum_i\!\Ab(\rb_i,t)\!\cdot\!\nabla_{\rb_i}, \\
\HH''(t)&\!=\frac{e^2Z^2}{2\mn c^2}A^2(\rn,t)+\frac{e^2}{2\me c^2}\sum_i A^2(\rb_i,t),
\end{align}
\end{subequations}
which are linear and quadratic in the classical vector potential $\Ab(\rb,t)$. We work in the Coulomb gauge ($\nabla_\rb\cdot\Ab(\rb,t)=0$), and due to the transverse character of electromagnetic fields in the vacuum region in which the atom is moving, the scalar potential is zero. Incidentally, we use the notation $A^2\equiv\Ab\cdot\Ab$ for vectors. Considering external fields that vary very little across the small size of the atom, we can approximate the vector potential at the nucleus and electron positions by the first-order Taylor expansion around the CM coordinates $\rb$ according to
\begin{align}
\Ab(\rn,t)&\approx\Ab(\rb,t)+(\ub\cdot\nabla_\rb)\,\Ab(\rb,t), \nonumber\\
\Ab(\rb_i,t)&\approx\Ab(\rb,t)+\big[(\ub_i+\ub)\cdot\nabla_\rb\big]\,\Ab(\rb,t).
\nonumber
\end{align}
Introducing these expressions in Eqs.~\eqref{Hint12}, and writing
\begin{subequations}
\label{nablatrans}
\begin{align}\nabla_{\rn}&=\frac{\mn}{M}\nabla_\rb-\sum_i\nabla_{\ub_i}, \\
\nabla_{\rb_i}&=\frac{\me}{M}\nabla_\rb+\nabla_{\ub_i}
\end{align}
\end{subequations}
from Eqs.~\eqref{rbub}, we find
\begin{subequations}
\label{Hint12bis}
\begin{align}
\HH'(t)\approx&\frac{\ii\hbar e}{\me c}\bigg\{-\frac{M}{\mn}\,\Ab(\rb,t)\cdot\sum_i\nabla_{\ub_i}
+\big[(\ub\cdot\nabla_\rb)\,\Ab(\rb,t)\big]\cdot\nabla_\rb\bigg\}, \label{Hint1}\\
\HH''(t)\approx&\frac{e^2}{2\me c^2}\frac{M}{\mn}\bigg\{Z\,A^2(\rb,t)
-\frac{1}{\me}(2\mn-M)\,\big[\ub\cdot\nabla_\rb\,A^2(\rb,t)\big]\bigg\}, \label{Hint2}
\end{align}
\end{subequations}
where we only retain terms linear in either $\ub_i$ or $\nabla_{\ub_i}$, as higher-order contributions produce corrections beyond the dipolar response of the atom, while the inhomogeneous term cancels exactly. For clarity, we note that the gradients inside $\big[(\ub\cdot\nabla_\rb)\,\Ab(\rb,t)\big]$ and $\big[\ub\cdot\nabla_\rb\,A^2(\rb,t)\big]$ only affect the vector field, whereas the rightmost operator $\nabla_\rb$ in Eq.~\eqref{Hint1} acts on the wave function in the Schr\"odinger equation (see below).

We now evaluate the matrix elements $\bra{j}\HH\ket{j'}$ by expressing them in terms of the transition dipoles $e\,\db_{jj'}$ with
\begin{align}
\db_{jj'}=Z\bra{j}\rn\ket{j'}-\!\sum_i\bra{j}\rb_i\ket{j'}=\!-\!\sum_i\bra{j}\ub_i\ket{j'}. \label{djj}
\end{align}
Using the identities
\begin{subequations}
\label{djjbis}
\begin{align}
\bra{j}\ub\ket{j'}&=\frac{\me}{M}\,\db_{jj'}, \label{djjbis1}\\
\sum_i\bra{j}\nabla_{\ub_i}\ket{j'}&=\frac{\mn\me}{\hbar M}\,\omega_{jj'}\,\db_{jj'} \label{djjbis2}
\end{align}
\end{subequations}
(see self-contained derivation below), where $\omega_{jj'}=\omega_j-\omega_{j'}$, together with Eqs.~\eqref{Hint12bis}, we find
\begin{subequations}
\label{Hjj}
\begin{align}
&\HH'_{jj'}(t)=\bra{j}\HH'(t)\ket{j'}\approx-\frac{\ii e}{c}\,\omega_{jj'}\,\db_{jj'}\cdot\Ab(\rb,t)
+\frac{\ii\hbar e}{Mc}\,\big[(\db_{jj'}\cdot\nabla_\rb)\,\Ab(\rb,t)\big]\cdot\nabla_\rb, \label{Hjj1}\\
&\HH''_{jj'}(t)=\bra{j}\HH'(t)\ket{j'}\approx\frac{e^2}{2\me c^2}\frac{M}{\mn}
\,\big[Z\,A^2(\rb,t)\,\delta_{jj'}-\big(2\mn/M-1\big)\,(\db_{jj'}\cdot\nabla_\rb)\,A^2(\rb,t)\big]. \label{Hjj2}
\end{align}
\end{subequations}
We note that the property $\db_{jj}=0$ is implicitly employed in the derivations that follow. Finally, matrix elements including translational degrees of freedom are readily obtained from Eqs.~\eqref{Hjj} by using CM momentum wave functions $\bra{\rb}\pb\rangle=\ee^{\ii\pb\cdot\rb}/\sqrt{V}$, where $V$ is the quantization volume. We find
\begin{subequations}
\label{Hjjfinal}
\begin{align}
&H'_{\pb j,\pb'j'}(t)=\bra{\pb j}\HH'(t)\ket{\pb'j'}
\approx-\frac{\ii e}{cV}\;\db_{jj'}\cdot\bigg\{\omega_{jj'}\,\Ab_{\pb-\pb'}(t)
+\frac{\hbar}{M}\,(\pb-\pb')\,\big[\pb'\cdot\Ab_{\pb-\pb'}(t)\big]\bigg\}, \label{Hjjfinal1}\\
&H''_{\pb j,\pb'j'}(t)=\bra{\pb j}\HH''(t)\ket{\pb'j'}
\approx\frac{e^2}{2\me c^2V}\frac{M}{\mn}\int\! d^3\rb \;\ee^{\ii(\pb'-\pb)\cdot\rb}\;
\bigg\{Z\,A^2(\rb,t)\,\delta_{jj'}-\ii\,\big(2\mn/M-1\big)\,[\db_{jj'}\cdot(\pb-\pb')]\,A^2(\rb,t)\bigg\}, \label{Hjjfinal2}
\end{align}
\end{subequations}
\end{widetext}
where $\Ab_{\qb}(t)$ is the vector potential in momentum space, defined by the Fourier transform
\begin{align}
\Ab_{\qb}(t)&=\int d^3\rb\;\ee^{-\ii\qb\cdot\rb}\,\Ab(\rb,t). \label{qspace}
\end{align}
In what follows, we also use frequency-space quantities defined by
\begin{align}
f_\omega&=\int dt\;\ee^{\ii\omega t}\,f(t) \label{wspace}
\end{align}
for any time-dependent function $f(t)$.

\section{Light-atom interaction: effective Hamiltonian}

The time-dependent state of the atom can be expanded in the complete set of free-atom states as
\begin{align}
\ket{\Psi(t)}&=\sum_{\pb j} c_{\pb j}(t)\,\ee^{-\ii(\varepsilon_\pb+\omega_j)t}\,\ket{\pb j}.
\nonumber
\end{align}
From the Schr\"odinger equation $\HH(t)\ket{\Psi(t)}=\ii\hbar\ket{\dot{\Psi}(t)}$, the expansion coefficients $c_{\pb j}(t)$ are found to follow the equation of motion
\begin{align}
\ii\hbar\,\dot{c}_{\pb j}(t)=\sum_{\pb'j'}&c_{\pb'j'}(t)\,\ee^{\ii(\varepsilon_{\pb\pb'}+\omega_{jj'}) t} \label{eqofm}\\
&\times\big[H'_{\pb j,\pb'j'}(t)+H''_{\pb j,\pb'j'}(t)\big], \nonumber
\end{align}
where $\varepsilon_{\pb\pb'}=\varepsilon_\pb-\varepsilon_{\pb'}$, whereas the matrix elements of $\HH'(t)$ and $\HH''(t)$ are given by Eqs.~\eqref{Hjjfinal}.

We assume the system to be initially prepared in the ground state $j=0$ [i.e., $c_{\pb j}(-\infty)=0$ for $j\neq0$], and further adopt the following two approximations: ($i$) the population of any excited state $\ket{j\neq0}$ is small compared to that of $\ket{0}$, so that we only consider transitions to $\ket{j\neq0}$ coming from $\ket{0}$; and ($ii$) the process is Markovian \cite{L1973}, implying that the probability of each event only depends on the state reached in the immediately previous event. We then follow a standard procedure consisting in solving $c_{\pb j}(t)$ for $j\neq0$ and introducing the result back in Eq.~\eqref{eqofm} to obtain an effective Schr\"odinger equation for the $j=0$ component. In doing so, we find that $\HH''$ produces a contribution $\propto\,A^2$ in $c_{\pb,j\neq0}(t)$, and thus terms $\propto\,A^4$ in the effective Hamiltonian, which are negligible for the field strengths considered in this work. Consequently, the solution for the expansion coefficients with $j\neq0$ can be approximated by integrating Eq.~\eqref{eqofm} without including $\HH''$. This leads to
\begin{align}
&c_{\pb j}(t)\approx-\frac{\ii}{\hbar}\sum_{\pb'}\int_{-\infty}^t dt'\;c_{\pb'0}(t')\,H'_{\pb j,\pb'0}(t')\,\ee^{\ii(\varepsilon_{\pb\pb'}+\omega_{j0}) t'} \nonumber\\
&\approx\!-\frac{1}{\hbar}\!\sum_{\pb'}c_{\pb'0}(t)\!\int\frac{d\omega}{2\pi}\,H'_{\pb j,\pb'0,\omega}\;\frac{\ee^{\ii(\varepsilon_{\pb\pb'}+\omega_{j0}-\omega) t}}{\varepsilon_{\pb\pb'}+\omega_{j0}-\omega-\ii\kappa_j/2},
\label{cpj}
\end{align}
where only $j=0$ terms are retained in the right-hand side in virtue of ($i$). The second line in this expression uses the frequency-space representation defined by Eq.~\eqref{wspace} and incorporates a phenomenological decay rate $\kappa_j$ of the excited atomic state $\ket{j}$. Also, we move $c_{\pb0}(t)$ outside the integral under the assumption that these coefficients vary slowly compared with the relatively fast oscillations of the exponential factor \cite{L1973} [approximation ($ii$)]. Then, we insert Eq.~\eqref{cpj} in the right-hand side of Eq.~\eqref{eqofm} and specify the left-hand side for $j=0$ to write
\begin{align}
\ii\hbar\,\dot{c}_{\pb0}(t)&=\sum_{\pb'}c_{\pb'0}(t)\,H^{\rm eff}_{\pb\pb'}(t)\,\ee^{\ii\varepsilon_{\pb\pb'}t},
\label{cpt}
\end{align}
where
\begin{align}
&H^{\rm eff}_{\pb\pb'}(t)=\frac{e^2Z}{2\me c^2V}\frac{M}{\mn}\int\! d^3\rb \;\ee^{\ii(\pb'-\pb)\cdot\rb}\,A^2(\rb,t) \label{Heff}\\
&-\frac{1}{\hbar}\sum_{\pb''j}\iint\frac{d\omega d\omega'}{(2\pi)^2}\,\ee^{-\ii(\omega+\omega')t}\,\frac{H'_{\pb0,\pb''j,\omega'}\,H'_{\pb'' j,\pb'0,\omega}}{\varepsilon_{\pb''\pb'}+\omega_{j0}-\omega-\ii\kappa_j/2}
\nonumber
\end{align}
is the effective interaction Hamiltonian that describes the evolution of the CM. We note that the first term in Eq.~\eqref{Heff} represents the only contribution remaining from $\HH''(t)$ at order $\propto\,A^2$ in the external field.

At this point, we adopt the nonrecoil approximation, so that the atom velocity $\vb$ is taken to be constant during the interaction with the optical field \cite{paper371}. This allows us to write
\begin{align}
\varepsilon_{\pb\pb'}\approx(\pb-\pb')\cdot\vb. \nonumber
\end{align}
In addition, we approximate $\pb'\approx M\vb/\hbar$ in the rightmost part of Eq.~\eqref{Hjjfinal1}, which becomes
\begin{align}
&H'_{\pb j,\pb'j',\omega} \label{Hpjpj}\\
&\approx-\frac{\ii e}{cV}\;\db_{jj'}\cdot\bigg\{\omega_{jj'}\,\Ab_{\pb-\pb',\omega}+(\pb-\pb')\,\big[\vb\cdot\Ab_{\pb-\pb',\omega}\big]\bigg\} \nonumber
\end{align}
in frequency space. We observe that $H'_{\pb j,\pb'j',\omega}=H'_{\pb-\pb',j,0j',\omega}$ is only a function of the wave vector difference $\pb-\pb'$ in the nonrecoil approximation, and consequently, $H^{\rm eff}_{\pb\pb'}(t)$ also depends on that difference alone. Using this property in Eq.~\eqref{cpt}, multiplying both sides by $\ee^{\ii\pb\cdot\rb-\ii\varepsilon_\pb t}$, and summing over $\pb$, we obtain the Schr\"odinger equation given in Eq.~\eqref{schr2}, where
\begin{align}
H^{\rm eff}(\rb,t)=\sum_\pb\ee^{\ii\pb\cdot\rb}\,H^{\rm eff}_{\pb0}(t)
\label{Heffrt1}
\end{align}
is the effective interaction Hamiltonian in real space,
\begin{align}
\psi(\rb,t)=V^{-1/2}\sum_\pb c_{\pb0}(t)\,\ee^{\ii\pb\cdot\rb-\ii\varepsilon_\pb t}
\nonumber
\end{align}
is the CM wave function of the atom, and we assume the latter to be tightly concentrated around central values of the kinetic energy $E_0$ and the wave vector $\qb_0$. In the derivation of Eq.~\eqref{schr2}, the kinetic part of the Hamiltonian is approximated as $\hbar\varepsilon_\qb=\hbar^2q^2/2M\approx E_0+\hbar(\qb-\qb_0)\cdot\vb$ in momentum space, followed by the substitution $\qb\rightarrow-\ii\nabla_\rb$ to move to real space.

\begin{widetext}
Substituting Eq.~\eqref{Hpjpj} in Eq.~\eqref{Heff}, and this in turn in Eq.~\eqref{Heffrt1}, we obtain
\begin{align}
&H^{\rm eff}(\rb,t)=\frac{e^2Z}{2\me c^2}\frac{M}{\mn}\,A^2(\rb,t) \label{Hefflast0}\\
&-\frac{e^2}{\hbar c^2}\sum_{j\neq0}\bigg\{\omega_{j0}\,\Ab(\rb,t)+\ii\nabla_\rb\,\big[\vb\cdot\Ab(\rb,t)\big]\bigg\} \nonumber\\
&\quad\cdot \int\frac{d^3\qb}{(2\pi)^3}\int\frac{d\omega}{2\pi}\,\ee^{\ii\qb\cdot\rb-\ii\omega t}\,\frac{\db_{0j}\otimes\db_{j0}}{(\omega_{j0}+\qb\cdot\vb-\omega-\ii\kappa_j/2)}\cdot\bigg\{\omega_{j0}\,\Ab_{\qb,\omega}+\qb\,\big[\vb\cdot\Ab_{\qb,\omega}\big]\bigg\},
\nonumber
\end{align}
where we have applied the customary prescription $\sum_\pb\rightarrow V\int d^3\pb/(2\pi)^3$ to transform momentum sums into integrals. Finally, considering isotropic atoms (or alternatively, averaging over orientations for magnetic atoms), the $j$ sum produces a diagonal tensor $\db_{0j}\otimes\db_{j0}$ proportional to the identity, so we can rewrite Eq.~\eqref{Hefflast0} as
\begin{align}
&H^{\rm eff}(\rb,t)=\frac{e^2Z}{2\me c^2}\frac{M}{\mn}\,A^2(\rb,t) \label{Hefflast}\\
&\quad\quad-\frac{e^2}{\hbar c^2}\sum_{j\neq0} \int\frac{d^3\qb}{(2\pi)^3}\int\frac{d\omega}{2\pi}\,\ee^{\ii\qb\cdot\rb-\ii\omega t}
\,\frac{|\xx\cdot\db_{j0}|^2}{(\omega_{j0}+\qb\cdot\vb-\omega-\ii\kappa_j/2)} \nonumber\\
&\quad\quad\quad\quad\quad\quad\quad\quad\times\bigg\{\omega_{j0}\,\Ab(\rb,t)+\ii\nabla_\rb\,\big[\vb\cdot\Ab(\rb,t)\big]\bigg\} \cdot\bigg\{\omega_{j0}\,\Ab_{\qb,\omega}+\qb\,\big[\vb\cdot\Ab_{\qb,\omega}\big]\bigg\},
\nonumber
\end{align}
which can be applied to any form of the classical optical field.

As a relevant configuration, we note that the effective Hamiltonian in Eq.~\eqref{Hefflast} can be simplified if the vector potential $\Ab(\rb,t)$ is perpendicular to the atom velocity $\vb$. Then, using the Thomas-Reiche-Kuhn (TRK) sum rule \cite{T1925,RT1925,K1925}
\begin{align}
\sum_{j\neq0} \omega_{j0}\,|\xx\cdot \db_{0j}|^2=\frac{\hbar Z}{2\me}\frac{M}{\mn}
\label{TRK}
\end{align}
(see self-contained derivation below), we can express the first term in the right-hand side of Eq.~\eqref{Hefflast} as a $j$ sum and write
\begin{align}
&H^{\rm eff}(\rb,t)=\frac{e^2}{\hbar c^2} \sum_{j\neq0} \omega_{j0}\,|\xx\cdot\db_{j0}|^2
\int\frac{d^3\qb}{(2\pi)^3}\int\frac{d\omega}{2\pi}
\,\ee^{\ii\qb\cdot\rb-\ii\omega t}
\,\frac{\qb\cdot\vb-\omega}{\omega_{j0}+\qb\cdot\vb-\omega-\ii\kappa_j/2} \,\Ab(\rb,t)\cdot\Ab_{\qb,\omega}. \label{Heff2}
\end{align}
\end{widetext}
This result is also valid for arbitrary $\Ab(\rb,t)$ orientation in the $v\ll c$ limit (see below). Finally, using the Fourier transforms in Eqs.~\eqref{qspace} and \eqref{wspace}, and neglecting $\kappa_j$, we obtain Eq.~\eqref{Heffsimplified} in the main text.

\section{PINAM interaction with monochromatic evanescent fields}

We consider an atom moving with velocity $\vb$ along the $z$ direction in the presence of an external optical electric field $\Eb(\rb,t)=\Eb(\Rb)\,\ee^{\ii q_zz-\ii\omega t}+{\rm c.c.}$ of well-defined frequency $\omega$ and longitudinal wave vector $q_z$, such that the vector potential can be chosen as $\Ab(\rb,t)=-(\ii c/\omega)\,\Eb(\Rb)\,\ee^{\ii q_zz-\ii\omega t}+{\rm c.c.}$ with $\Rb=(x,y)$. Inserting this expression in Eq.~\eqref{Hefflast}, we find that the integrand in Eq.~\eqref{solution} is composed of a term independent of $t'$ and two more terms proportional to $\exp\{\pm2\ii (q_zz-\omega)t'\}$. The $t'$-independent term produces an elastic phase, whereas the other two give rise to the sort of inelastic events in which we are interested here. When the interaction region extends over a large length $L$ along the trajectory, those two terms contribute negligibly to the integral, unless the phase-matching condition $\omega=q_zv$ [i.e., Eq.~\eqref{phasematching1}] is satisfied, as we assume here. Then, the integral over $t'$ simply yields a factor $L/v$ (the interaction time) and the post-interaction CM wave function reduces to
\begin{widetext}
\begin{align}
&\psi(\rb,t)=\psi^{\rm inc}(\rb,t)\,\ee^{\ii\varphi(\Rb)}
\,\exp\big\{-\beta(\Rb)\,\ee^{2\ii\omega(z-vt)/v}+\beta^*(\Rb)\,\ee^{-2\ii\omega(z-vt)/v}\big\},
\label{psimono1}
\end{align}
where
\begin{subequations}
\label{phibeta1}
\begin{align}
\varphi(\Rb)&=-\frac{e^2ZL}{\hbar\me v\omega^2}\frac{M}{\mn}\,\big|\Eb(\Rb)\big|^2+\frac{2e^2L}{\hbar^2v\omega^2} \sum_{j\neq0} \frac{|\xx\cdot \db_{0j}|^2}{\omega_{j0}}
\bigg(\omega^2_{j0}\,\big|\Eb(\Rb)\big|^2+v^2\big|(q_z\zz-\ii\nabla_\Rb)\,E_z(\Rb)\big|^2\bigg), \\\beta(\Rb)&=-\frac{\ii e^2ZL}{2\hbar\me v\omega^2}\frac{M}{\mn}\,E^2(\Rb)
+\frac{\ii e^2L}{\hbar^2v\omega^2} \sum_{j\neq0} \frac{|\xx\cdot \db_{0j}|^2}{\omega_{j0}}
\bigg(\omega^2_{j0}\,E^2(\Rb)-v^2\big[(q_z\zz-\ii\nabla_\Rb)\,E_z(\Rb)\big]^2\bigg)
\end{align}
\end{subequations}
\end{widetext}
are an elastic phase and an inelastic atom-light coupling coefficient, respectively, and we have neglected $\kappa_j$ in front of $\omega_{j0}$. This result can be simplified by using the TRK sum rule \cite{T1925,RT1925,K1925} given in Eq.~\eqref{TRK}, so that the first term inside the $j$ sum exactly cancels the first term in the right-hand side of Eqs.~\eqref{phibeta1}. Then, we have
\begin{subequations}
\label{phibeta2}
\begin{align}
&\varphi(\Rb)=\frac{L}{\hbar v}\,\alpha_0\,\big|(\zz-\ii(v/\omega)\nabla_\Rb)\,E_z(\Rb)\big|^2, \\
&\beta(\Rb)=-\frac{\ii L}{2\hbar v}\,\alpha_0\,\big[(\zz-\ii (v/\omega)\nabla_\Rb)\,E_z(\Rb)\big]^2,
\end{align}
\end{subequations}
where $\alpha_0=(2e^2/\hbar)\sum_{j\neq0}|\xx\cdot\db_{0j}|^2/\omega_{j0}$ is the electrostatic polarizability of the atom \cite{BJ03}.

The coefficient $\beta(\Rb)$ plays a role similar to the coupling coefficient in photon-induced near-field electron microscopy \cite{BFZ09} (PINEM), but in contrast to a linear dependence on the field, we find a quadratic scaling. Following similar methods as in PINEM \cite{PLZ10,paper371}, we now use the Jacobi-Anger formula to recast Eq.~\eqref{psimono1} into Eq.~\eqref{psigeneral} with $\Omega=2\omega$. This expression establishes the basis for PINAM. Each $\ell$ term in this expansion represents a wave function component that is displaced in momentum by $2\ell\hbar\omega/v$ and in energy by $2\ell\hbar\omega$, therefore spanning a series of sidebands with a probability density $J^2_\ell(2|\beta(\Rb)|)$ that depends on the transverse coordinates $\Rb$.

As a possible realization of phase-matching monochromatic illumination, we consider an atom reflected from a polariton-supporting planar surface under grazing incidence conditions. To analyze the coupling strength, we then assume parallel motion of the atom relative to the surface, as well as a polariton characterized by an in-plane wave vector $q_z$ and frequency $\omega$ satisfying the phase-matching condition in Eq.~\eqref{phasematching1}. In the vacuum region outside the surface, the electric field associated with the polariton takes the general form $\mathcal{E}_0(\zz+\ii\gamma\,\xx)\,\ee^{\ii q_zz-\kappa_px-\ii\omega t}+{\rm c.c.}$, where $z$ and $x$ are parallel and perpendicular to the surface, $\gamma=1/\sqrt{1-v^2/c^2}$ is the relativistic Lorentz factor, $\kappa_p=\omega/v\gamma$ is taken to be real (i.e., we neglect inelastic losses of the surface mode), and $\mathcal{E}_0$ is a global field amplitude. We thus have $E_z(\Rb)=\mathcal{E}_0\ee^{-k_px}$, for which Eqs.~\eqref{phibeta2} become
\begin{align}
&\varphi(\Rb)=\frac{L}{\hbar v}\,\alpha_0|\mathcal{E}_0|^2\,\ee^{-2\kappa_px}(2-v^2/c^2), \nonumber\\
&\beta(\Rb)=-\frac{\ii vL}{2\hbar c^2}\,\alpha_0\mathcal{E}_0^2\,\ee^{-2\kappa_px}. \nonumber
\end{align}
In this geometry, the transverse component partially cancels the longitudinal one in $(-\ii v\nabla_\Rb+\omega\zz)\,E_z(\Rb)$, resulting in a $\beta(\Rb)\propto v/c^2$ scaling with atom velocity. This cancellation is dramatic for $v\ll c$.

A more advantageous situation in which such cancellation does not take place is encountered near curved surfaces and also in a vacuum region flanked by two planar surfaces, for which the transverse derivative of $E_z(\Rb)$ vanishes at some positions and we have a $\beta(\Rb)\propto1/v$ scaling (see below). As an arrangement of practical interest, we consider a circular hole of radius $a$ running parallel to $z$ and drilled in a homogeneous polaritonic material of permittivity $\epsilon(\omega)$. This configuration could be realized by perforating holes in a film of large thickness $L$, so that an atom beam impinging normally to the film can be exposed to optically driven polaritons as it moves along the holes. For $\omega a/c\ll1$, we can operate in the electrostatic limit, so that any propagating polariton confined to the hole can be described by means of a potential $\phi(\rb,t)=A I_m(q_z R)\ee^{\ii q_zz+\ii m\varphi-\ii\omega t}+{\rm c.c.}$ in the interior region ($R<a$) and $\phi(\rb,t)=B K_m(q_z R)\ee^{\ii q_zz+\ii m\varphi-\ii\omega t}+{\rm c.c.}$ outside it ($R>a$). Here, $m$ is the azimuthal number, we use cylindrical coordinates $\rb=(R,\varphi,z)$, $K_m$ and $I_m$ are modified Bessel functions, and the constants $A$ and $B$ are subject to the conditions $B/A=I_m(q_za)/K_m(q_za)=\epsilon(\omega)I'_m(q_za)/K'_m(q_za)$, which guarantee the continuity of the potential and the normal electric displacement at the hole surface. We take $m=0$, for which the $z$ component of the electric field amplitude reduces to $E_z(\Rb)=\mathcal{E}_0 I_0(q_zR)$, subject to the condition (after using the Wronskian \cite{AS1972}) $[1-\epsilon(\omega)]\,q_za\,I_1(q_za)K_0(q_za)=1$ for the existence of a polariton with longitudinal wave vector $q_z$ and frequency $\omega$. Inserting this field inside Eqs.~\eqref{phibeta2}, where the phase-matching condition in Eq.~\eqref{phasematching1} is assumed, we obtain
\begin{align}
&\varphi(\Rb)=\frac{L}{\hbar v}\,\alpha_0|\mathcal{E}_0|^2\,\big[I_0^2(\omega R/v)+I_1^2(\omega R/v)\big], \nonumber\\
&\beta(\Rb)=-\frac{\ii L}{2\hbar v}\,\alpha_0\mathcal{E}_0^2\,\big[I_0^2(\omega R/v)-I_1^2(\omega R/v)\big], \nonumber
\end{align}
and in particular, for an atom moving along the hole axis ($R=0$), we find that $\beta$ reduces to Eq.~\eqref{betamaintext} in the main text. Note that $\beta(\Rb)$ grows monotonically for small $\omega R/v$ and takes values close to that in Eq.~\eqref{betamaintext} up to $\omega R/v\sim1$ for $m=0$, so this equation is a good approximation also for off-axis trajectories and relatively small holes of radius $a\lesssim v/\omega$.

\section{Stimulated Compton scattering under two-color illumination}

We now focus on slow atoms ($v\ll c$) moving in free space in the presence of propagating optical fields composed of frequencies comparable to or smaller than the excitation frequencies, so that $qv$ can be neglected in front of $\omega$ and $\omega_{j0}$. Under these conditions, the effective Hamiltonian in Eq.~\eqref{Hefflast} can be approximated by neglecting the $\vb\cdot\Ab$ terms and applying the TRK sum rule \cite{T1925,RT1925,K1925}. This leads to Eq.~\eqref{Heff2}, which we use in what follows as the effective Hamiltonian.

Although inelastic CM scattering is kinematically forbidden for monochromatic fields in free space, it is possible to conserve energy and momentum via Compton scattering using two-color light. We consider illumination with two plane waves $i=1,2$ of wave vectors $\qb_i$ and frequencies $\omega_i$, such that the optical electric field is $\Eb(\rb,t)=\sum_{i=1,2}\Eb_i\,\ee^{\ii\qb_i\cdot\rb-\ii\omega_it}+{\rm c.c.}$, while the vector potential reduces to $\Ab(\rb,t)=-\sum_{i=1,2}(\ii c/\omega_i)\,\Eb_i\,\ee^{\ii\qb_i\cdot\rb-\ii\omega_it}+{\rm c.c.}$ in the Coulomb gauge. Inserting this expression into Eq.~\eqref{Heff2}, and this in turn into Eq.~\eqref{solution}, we find that the integrand in the latter contains terms that are (1) independent of $t'$, (2) proportional to $\exp\big\{\!\!\pm\ii\big[(\qb_1\cdot\vb-\omega_1)-(\qb_2\cdot\vb-\omega_2)\big]t'\big\}$, (3) proportional to $\exp\big\{\!\!\pm\ii\big[(\qb_1\cdot\vb-\omega_1)+(\qb_2\cdot\vb-\omega_2)\big]t'\big\}$, or (4) proportional to $\exp\big\{\!\!\pm2\ii(\qb_i\cdot\vb-\omega_i)t'\big\}$ with $i=1,2$. Terms of types (3) and (4) vanish after integrating over a large interaction time $L/v$ because of the inequality $|\qb_i\cdot\vb|\le\omega_iv/c<\omega_i$ for propagating waves, so the exponents take nonzero values, giving rise to self-compensating oscillations. The remaining terms produce (1) a global elastic phase $\varphi$ and (2) inelastic transitions described by a coupling coefficient $\beta$, subject to the phase-matching condition in Eq.~\eqref{phasematching2}, which is general for Compton scattering. We assume this condition to hold for our two incident plane waves. After some tedious but straightforward algebra, the post-interaction CM wave function becomes
\begin{align}
&\psi(\rb,t)=\psi^{\rm inc}(\rb,t)\,\ee^{\ii\varphi}\nonumber\\
&\quad\times\exp\big\{-\beta\,\ee^{\ii(\qb_1-\qb_2)\cdot(\rb-\vb t)}+\beta^*\,\ee^{-\ii(\qb_1-\qb_2)\cdot(\rb-\vb t)}\big\},
\nonumber
\end{align}
or equivalently, using the Jacobi-Anger formula \cite{PLZ10,paper371} and taking $\vb\parallel\zz$, we recover Eq.~\eqref{psigeneral} with $\Omega=\omega_1-\omega_2$, a global phase
\begin{align}
\varphi&=\frac{2e^2L}{\hbar^2v}\!\sum_{i=1,2}\!\big|\Eb_i(\Rb)\big|^2\sum_{j\neq0}
\frac{\omega_{j0}\,|\xx\cdot \db_{0j}|^2}{\omega_{j0}^2-(\omega_1-\qb_1\cdot\vb)^2},
\label{phiCompton}
\end{align}
and a coupling coefficient given by Eq.~\eqref{betaCompton}. In the derivation of Eqs.~\eqref{betaCompton} and \eqref{phiCompton}, we have again considered $v\ll c$, except in the rightmost denominators, in which $\qb_1\cdot\vb$ is maintained because it can be comparable to $\omega_{j0}-\omega_1$ for near-resonance illumination.

\section{Derivation of equations~\eqref{djjbis}}

We find Eq.~\eqref{djjbis1} directly from Eq.~\eqref{djj} by using the relation $\ub=(-\me/M)\sum_i\ub_i$. For the derivation of Eq.~\eqref{djjbis2}, we start from the free-atom Hamiltonian $\HH^{\rm at}=-(\hbar^2/2\mn)\nabla^2_{\rn}-(\hbar^2/2\me)\sum_i\nabla^2_{\rb_i}+\hat{V}$, where $\hat{V}$ accounts for the electron-nucleus and electron-electron Coulomb interactions. Using Eqs.~\eqref{nablatrans}, we can rewrite this operator as $\HH^{\rm at}=-(\hbar^2/2M)\nabla^2_{\rb}-(\hbar^2/2\me)\sum_i\nabla^2_{\ub_i}-(\hbar^2/2\mn)\sum_{ii'}\nabla_{\ub_i}\cdot\nabla_{\ub_{i'}}+\hat{V}$. Upon direct inspection, we find the commutator $\big[\ub_i,\HH^{\rm at}\big]=(\hbar^2/\me)\nabla_{\ub_i}+(\hbar^2/\mn)\sum_{i'}\nabla_{\ub_{i'}}$, where the second term is common for all $\ub_i$'s. Taking the matrix elements of the right- and left-hand sides of this identity, we find $\hbar(\omega_{j'}-\omega_j)\bra{j}\ub_i\ket{j'}=(\hbar^2/\me)\bra{j}\nabla_{\ub_i}\ket{j'}+(\hbar^2/\mn)\bra{j}\sum_i\nabla_{\ub_i}\ket{j'}$. Finally, summing over $i$, noticing that we have $Z$ electrons (i.e., the atom is neutral), and using Eq.~\eqref{djjbis1}, we readily obtain Eq.~\eqref{djjbis2}.

\section{Derivation of equation~\eqref{TRK}}

We start from the TRK sum rule \cite{T1925,RT1925,K1925} $\bra{0}[\hat{A},[\HH,\hat{A}]]\ket{0}=2\hbar\sum_j\omega_{j0}\big|\bra{j}\hat{A}\ket{0}\big|^2$, which is valid for any Hermitian operator $\hat{A}$ and Hamiltonian $\HH$. Applying it to $\hat{A}=\sum_i\xx\cdot\ub_i\equiv\sum_iu_{ix}$ and $\HH^{\rm at}$, and making use of the commutator $\big[u_{ix},\HH^{\rm at}\big]=(\hbar^2/\me)\partial_{u_{ix}}+(\hbar^2/\mn)\sum_{i'}\partial_{u_{i'x}}$ derived in the previous paragraph together with the definition of $\db_{j0}$ in Eq.~\eqref{djj}, we directly obtain Eq.~\eqref{TRK}.



\end{document}